# Gamma-ray bursts as cool synchrotron sources

J. Michael Burgess[1,2], Damien Bégué[1], Ana Bacelj[1,3], Dimitrios Giannios[4], Francesco Berlato[1,5], and Jochen Greiner[1,2]

[1] Max-Planck Institut für Extraterrestrische Physik, Giessenbachstr. 1, 85748 Garching, Germany
[2] Excellence Cluster Universe, Technische Universität München, Boltzmannstr. 2, 85748, Garching, Germany
[3] Department of Physics, University of Rijeka, 51000 Rijeka, Croatia
[4] Department of Physics and Astronomy, Purdue University, 525 Northwestern Avenue, West Lafayette, IN 47907, USA
[5] Physik Department, Technische Universität München, D-85748 Garching, James-Franck-Strasse 1, Germany

**Gamma-ray bursts are the most energetic electromagnetic sources in the Universe. Their prompt gamma-ray radiation, lasting between a fraction of a second to several thousand seconds, corresponds to an energy release of $10^{42}$-$10^{47}$ J [1,2]. Fifty years after their discovery and several dedicated space-based instruments, the physical origin of this emission is still unknown. Synchrotron emission has been one of the early contenders[3,4], but was criticized because spectral fits of empirical models (such as a smoothly-connected broken power law or a cut-off power law) suggest too hard a slope of the low-energy power law, violating the so-called synchrotron line-of-death[5,6], reviving models of photospheric emission[7-9]. Fitting proper synchrotron spectra[10-12] (rather than heuristic functions) was first shown to work for individual GRBs[13,14], though without tracking electron cooling. When the latter was taken into account, several GRB spectra could be fit successfully[10]. Here we show that idealized synchrotron emission, when properly incorporating time-dependent cooling of the electrons, is capable of fitting ~95% of all time-resolved spectra of single-peaked GRBs as measured with Fermi/GBM. The comparison with spectral fit results based on previous empirical models demonstrates that the past exclusion of synchrotron radiation as an emission mechanism derived via the line-of-death was misleading. Our analysis probes the physics of these ultra-relativistic outflows and the microphysical processes which cause them to shine, and for the first time provides estimates of magnetic field strength and Lorentz factors of the emitting region directly**

**from spectral fits. The parameter distributions that we find are grossly compatible with theoretical spectral[15-17] and outflow[18] predictions. The emission energetics remain challenging for all theoretical models. As synchrotron radiation alone can explain the observed emission, it is difficult to reconcile the time scales, efficiencies, and microphysics predicted by relativistic Fermi shock acceleration[19] and the fireball model[20] with the observations. Thus, our modeling of the Fermi/GBM observations provides evidence that GRBs are produced by moderately magnetized jets in which relativistic mini-jets emit optically-thin synchrotron radiation at large emission radii.**

We perform time-resolved spectral analysis of the prompt spectra of gamma-ray bursts (GRBs) detected with the Fermi Gamma-ray Burst Monitor (GBM). We select a subset of GRBs which exhibit a single contiguous, pulse-like structure which is justified by the assumption that the emission originates from a single dissipation episode[10]. In addition, we require that all GRBs have a measured redshift to ascertain their energetics. This results in a sample of 19 out of the 81 GRBs in the GBM time-resolved spectral catalog[21]. After background modeling, we employ the Bayesian blocks algorithm to create spectral datasets in optimised temporal bins for each of those GRBs, leading to a total of 162 bins, each with a Poisson-Gaussian significance larger than $5\sigma$ which are then used for spectral fitting (see Methods, 'Observation and data analysis').

We describe the GRB prompt emission from a fully physical, time-dependent synchrotron spectral model. The synchrotron emission is modeled by assuming a generic electron acceleration mechanism that continuously injects a power law of electrons in the form of $Q(\gamma) \propto \gamma^{-p}$ ; $\gamma_{inj} \leq \gamma \leq \gamma_{max}$ . The electrons are subject to synchrotron cooling via a magnetic field and cool to a characteristic energy, $\gamma_{cool}$, during the emission period (see Methods, 'Synchrotron Modeling'). The emitted synchrotron radiation is computed during this cooling and summed to produce the model photon number spectrum (Fig. 1). This spectrum is then forward-folded through the instrument response and fitted to the GBM data via Bayesian posterior simulation. Each time-resolved spectrum is individually fitted and the data conditioned models are criticized using posterior predictive checks (PPCs). From this check, only 8 fits were inadequate and discarded from further analysis (see Methods, 'Model checking'). The high percentage of well-fit spectra is surprising, given that in the canonical approach[21] only 9.3% spectra are well-fit with a smoothly-connected broken power law, the so-called Band function[22], 11.4% by a smoothly-broken power law, 69.1% by an exponentially cutoff power law and 10.2% by a power law spectrum, all with low-energy slopes that often violate the line-of-death.

The success of these spectral fits to single-pulse GRBs overall is a confirmation of a synchrotron interpretation of prompt GRB emission, once time-dependence and cooling are properly included. They reveal that the electron distributions in GRB outflows range from extremely cooled to extremely uncooled by the emission of synchrotron photons (Fig. 2). The synchrotron cooling regime is typically defined as the ratio of $\chi = \gamma_{cool}/\gamma_{inj}$. The emission is said to be slow-cooling if $\chi > 1$ and fast-cooling if $\chi < 1$ [23]. We find both slow and fast cooling synchrotron emission within single GRBs (Fig. 3), with evidence of evolution of the cooling regime (Extended Data Fig. 1).

How were such results missed in previous analyses? Fitting of the measured data with the empirical Band function has lead to erroneous physical conclusions about the nature of GRB emission spectra. We demonstrate this by also fitting the Band function to the same spectra and show that the low-energy spectral slope (α) does not correlate to the cooling regime and as such to the proper physical model (Fig. 4). Historically, the spectral slope α has been used to infer or reject the physical emission mechanisms generating the observed prompt radiation, arguing that when $\alpha \geq -2/3$ (the asymptotic limit of synchrotron emission, corresponding to the so-called line-of-death), then the fitted spectrum cannot be produced by synchrotron emission. Comparing our two fits with the synchrotron model and the Band function, respectively, demonstrates that this argument does not hold as data with a fitted spectral slope $-2/3 \leq \alpha \leq 1$ via the Band function can be adequately fit with the synchrotron model. Nearly all spectral shapes reported in past GRB spectral analysis can be explained as synchrotron emission. An inferred thermal emission from hard spectra, i.e., with α~1 appears now as false interpretation. Our results do not explicitly rule out photospheric emission as the dominant source of GRB prompt spectra, but we have been able to overcome the main motivation for invoking photospheric emission or extra photospheric components, i.e., the hard values of α. To date, only GRB 090902B shows a clear sign of a photospheric origin[24].

On a more physical notion, our spectral fitting results have several implications:
(1) The fact that not all spectra are in the fast-cooling regime, challenges models of relativistic shock acceleration as their inherent inefficiency for dissipating internal energy into accelerated electrons requires the extraction of the maximum amount of energy from these electrons via synchrotron cooling[25]. Using our fitted parameters and assuming an emission radius of R=$10^{14}$ cm as predicted by models invoking shocks within the outflow, we can estimate the number of emitting electrons in each GRB, finding values of $\sim 10^{51} < N_e < 10^{56}$, consistent with previous findings (Extended Data Fig. 2) as well as magnetic field strengths of $\sim 10^{-2} G < B < 10^2$ G (Extended Data Fig. 3). We further find that the electrons are injected with a spectral index of p ~ 3.5 (Extended Data Fig. 4) which is different from the canonical relativistic shock prediction of p=2.24 [19]. In the interpretation of a shock origin of electron acceleration, this would be a sign of super-luminal shock geometry[26].

(2) GRBs have often been interpreted in the framework of the fireball model[20], which assumes that a large amount of thermal energy is released during the collapse of a massive star by a central engine such as an accreting black hole or a rapidly spinning neutron star. Under thermal pressure, the plasma accelerates to highly relativistic speeds by converting its thermal energy to bulk kinetic energy. When the outflow becomes optically thin, the remaining thermal energy is released in the form of photons producing a photospheric emission component with a quasi-thermal energy distribution[24]. In our analysis, we need no photospheric component in the spectra, however, since our simplistic synchrotron model already fits the vast majority of spectra. This imposes an upper limit on the bulk Lorentz factor of the outflow. Assuming a radiative efficiency of 10% and an initial radius of expansion of $10^7$ cm leads to bulk Lorentz factors on the order of a few tens for most GRBs analyzed herein. Those values are much smaller than previous results based on the observation of very high energy photons. A lower limit on bulk Lorentz factors can be obtained through an estimate of the opacity of the outflow to pair production[27]. We find that the two limits are incompatible ("Methods", Physical Calculations; Extended Data Fig. 5). This is a clear signature of magnetized jets, which can possess a large Lorentz factor without a strong thermal component[28].

(3) Instead of assuming other emission radii and checking for consistency, we can parameterize the problem as a function of the fractional magnetization ($\xi_B$). This leads to a set of equations for the Lorentz factor $\Gamma$, radius R, total particle number N, comoving magnetic field B and injection Lorentz factor $\gamma_{inj}$ (see Methods, "Physical Calculations"), from which we can infer radii for the entire sample assuming values for the magnetization from $10^{-4} - 1$. Furthermore, additional comoving bulk motion of blobs or minijet, with Lorentz factor $\Gamma_{em}$, is a possibility in models of magnetized dissipation. We find that moderate magnetization with $\xi_B \sim 1$ requires $\Gamma_{em} > 10$ in order to have the emission radius smaller than the radius at which the jet begins to decelerate at $\sim 10^{17}$ cm (see Extended Data Fig. 6 and 7). This is fully within the predictions[17,18]. Thus, our observations of synchrotron emission requires a large emission radius.

(4) A fourth implication is that other physical synchrotron models like, e.g. one which incorporates a temporally decaying magnetic field and increasing electron number injection[11], are not required: though a plausible physical scenario, we find that we do not need this feature to describe the data as we can model the spectra simply by allowing for the electrons to cool with time. Implementing a radially (and thus temporally) decaying magnetic field, as possible in a relativistically expanding blast wave of a GRB, provides curved low-energy spectra below the spectral peak energy most of which are consistent with the Band function[15]. This was shown to successfully fit only two GRBs, GRB 130606B[11] and GRB 160625B[12] though it suggested that the line-of-death is not a hard limit for the synchrotron model[11]. Moreover, these model fits have the problem that the radially decaying magnetization requires a rapidly increasing injection of electrons from an unspecified reservoir. (see Extended Data Fig. 8-9).

(5) Finally, as mentioned already earlier, we find a soft electron injection spectral index of p ~ 3.5. If the acceleration mechanism is magnetic reconnection, then the index of the electrons is directly tied to the level of magnetization in the jet according to PIC simulations[29]. The soft electron indices we observe are in line with the requirement of moderate magnetization. This presents a challenge to be tackled via PIC simulations with high magnetization, a yet unexplored parameter space, as our observations disagree with current speculative predictions. The lack of cooling observed in some spectra may be a direct signature of reheating of electrons via second-order Fermi acceleration[17,30].

The ability to model GRB spectra consistently as synchrotron emission is comforting as the model's simplicity is a convincing advance. The natural next steps are (i) finding physically meaningful dissipation mechanisms consistent with our simplistic synchrotron description, and (ii) reconciling the newly modeled observed spectral emission with the extreme energetics of these events. Ultimately, the implications of our model demand an exploration of both dynamics and particle acceleration in highly magnetized astrophysical outflows.

**Acknowledgements.**

We thank Johannes Buchner, Ewan Cameron and Roland Diehl for enlightening conversations. JMB and JG acknowledge support by the DFG cluster of excellence "Origin and Structure of the Universe" (www.universe-cluster.de). DB, FB and JG acknowledge support by the DFG through SFB 1258. AB acknowledges support by University of Rijeka through the Erasmus+ programme.

**Author Contributions.**

JMB has performed the writing, numerical modeling, and spectral analysis of the data. DB and DG provided theoretical calculations, interpretation and insight. AB performed sample selection and data reduction. JG and FB assisted in discussion and writing.

**Figure 1: Relation between electrons and emission spectrum.**
A schematic of the electron cooling and emitted photon spectra from our radiative code. Three regimes are displayed from dark red to pink: slow-cooling, moderate cooling and fast cooling. In the slow-cooling regime, the so-called cooling break is slight and barely noticeable. Note that the peak energy does not always correspond to the cooling break, in contrast with usual assumptions[26].

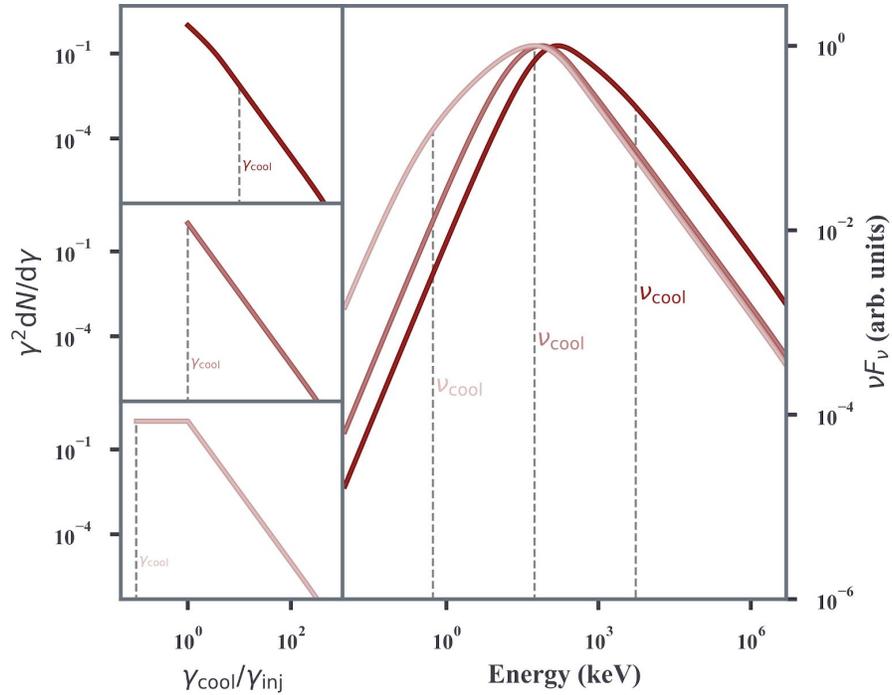

**Figure 2: From data to the electrons**

Fits to three time bins from GRB 130518580 demonstrate the diversity of cooling regimes encountered in GRBs. Our approach is clearly displayed from top to bottom where the data are fitted and overplotted with the posterior of the model for each detector in count space, followed by the observed $\nu F_\nu$ spectrum in model space with the marginal distributions of the characteristic frequencies overplotted. Finally, the posterior of the modeled electron spectrum is displayed where the color of each trace indicates how cool the electrons are, moving from slow-cooled (blue) to moderate cooling (green) and fast cooled (red).

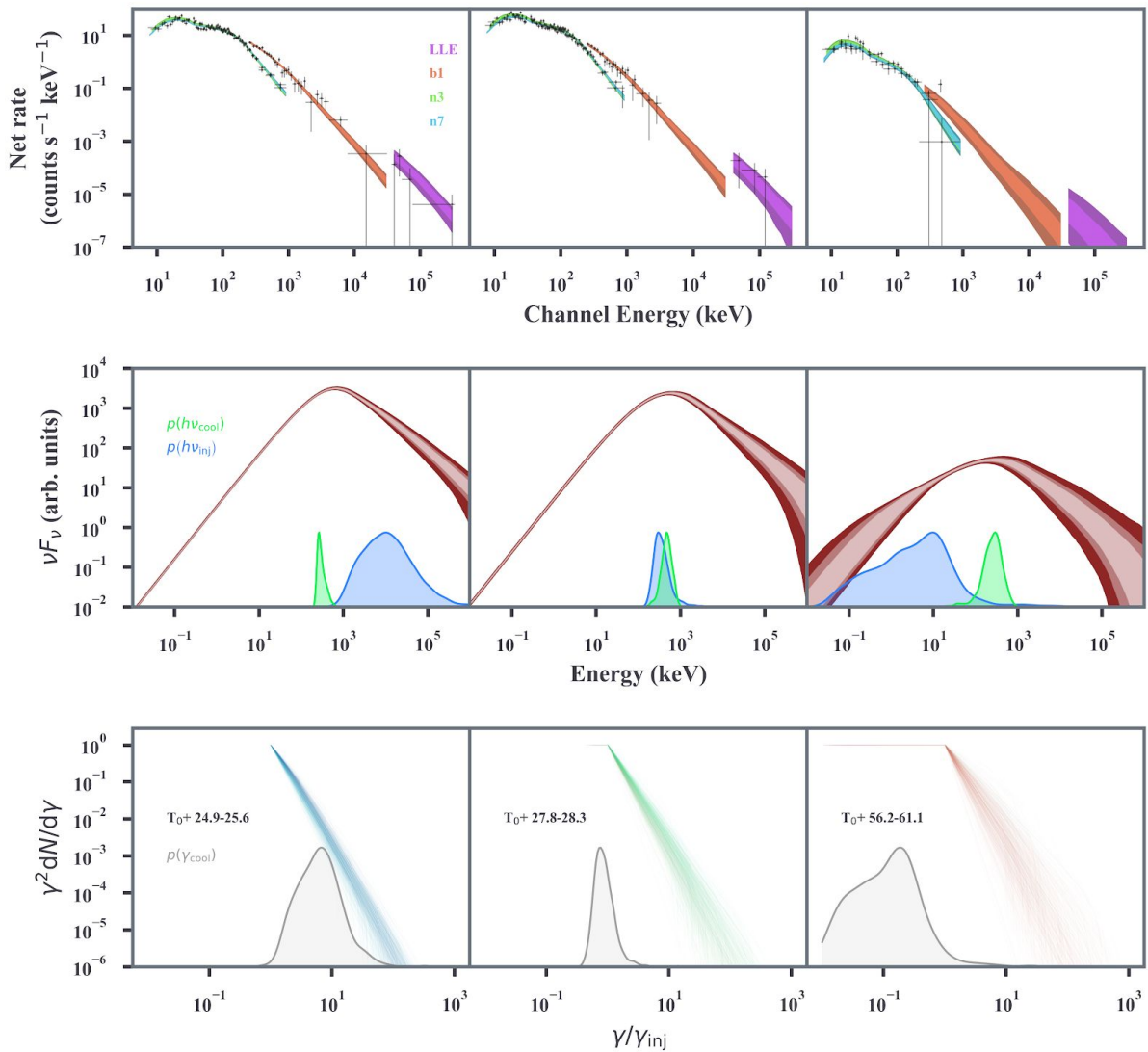

**Figure 3: Synchrotron emission can be cool**

The marginal distributions of $\chi = \gamma_{cool}/\gamma_{inj}$ for each well-fitted time-interval (individual red lines) of each GRB (separate panels). The grey region indicates the fast-cooling regime. The variety of cooling regimes observed is indicative of a plethora of different physical conditions occurring in GRB outflows.

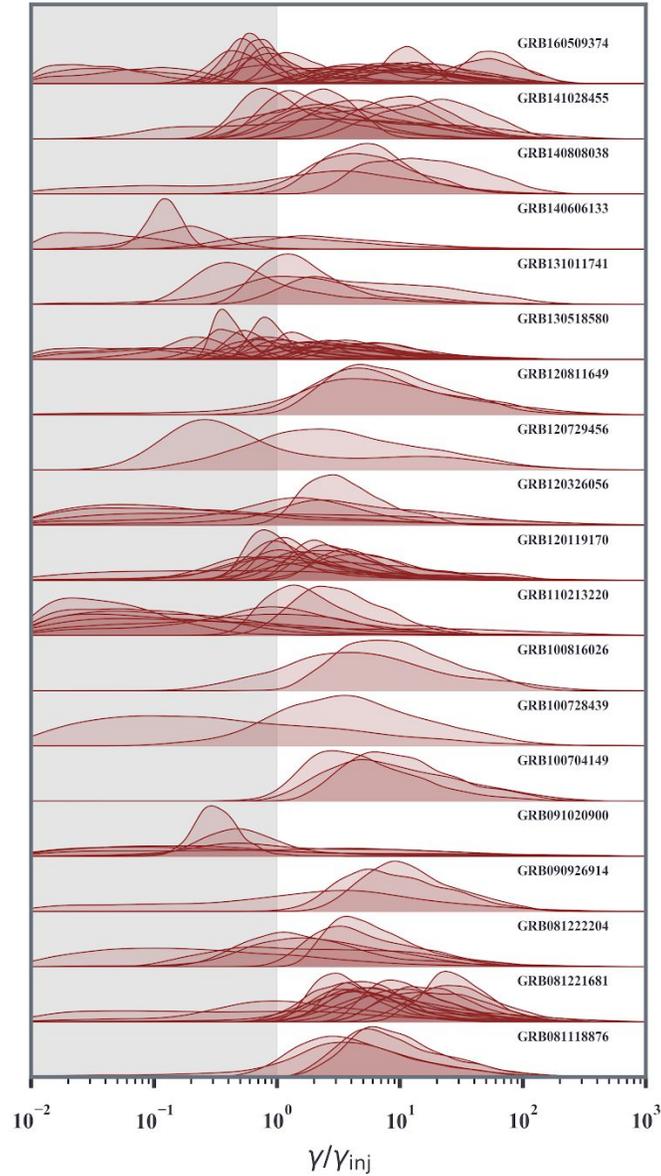

**Figure 4: Synchrotron defies death**

Marginal distributions of the Band function's low-energy slope (α) from fitted spectra are displayed and ranked according to the median cooling regime of the synchrotron from the same spectrum. The cooling regime loosely follows the asymptotic predictions of the line-of-death, however, several spectra well fitted by synchrotron have alpha values that strongly violate the line of death, i.e, α > -2/3.

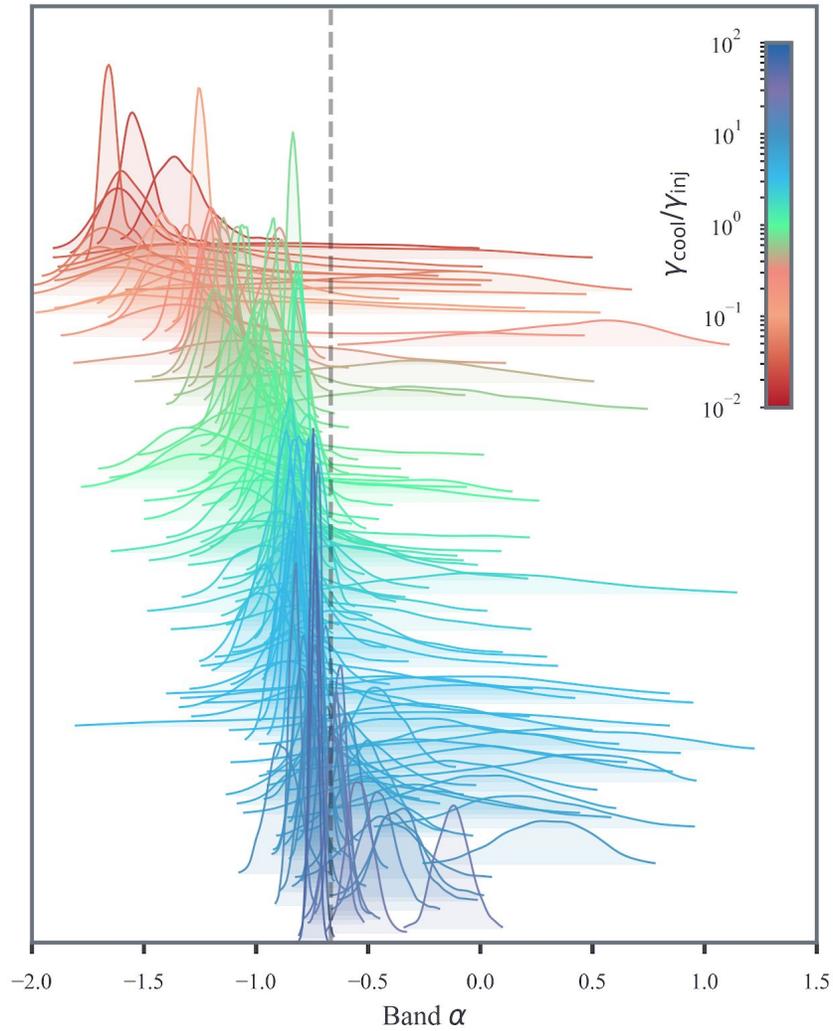

# Methods

**Observations and data analysis**

For each GRB in our sample, we select both NaI and BGO detectors with source viewing angles less than 60°. In the case of three GRBs, we include Fermi/ LAT LLE data. The background in all data is modeled with a polynomial in time fitted via an unbinned Poisson likelihood. With the background fitted, time bins are created via the Bayesian blocks algorithm[31] with a chance probability of $p_0$=0.05. Finally, only time bins with a Poisson-Gaussian[32] significance of 5σ or greater are selected for spectral analysis.

Bayesian posterior simulation is used to infer the spectral parameters of the model conditioned on the data. The appropriate likelihood for GBM and LLE data is the Poisson-Gaussian likelihood which accounts for the Poisson distribution of the total counts given the Gaussian distributed expected background from the polynomial temporal modeling. The posterior is sampled via the MULTINEST algorithm[33] because it can easily handle the multi-modal posterior present in some of the spectral parameter distributions. As we do not attempt model comparison, we do not use the marginal likelihood integral calculated via MULTINEST. Each spectral fit is performed with 500 live points.

The priors assumed for the parameters are a mixture of informative and uninformative distributions. For the normalization (K) we choose an uninformative log-uniform prior. Similarly for $B$, a scale parameter, a log-uniform prior is utilized such that the combination of $B$ and $\gamma_{inj}$ results in a spectral peak within the Fermi energy window. The prior for $\gamma_{cool}$ is a log-uniform distribution bounded by $10^{-2}\gamma_{inj}$ and $\gamma_{max}$. Lower values of $\gamma_{cool}$ would produce spectral features unidentified in the Fermi energy window and are computationally expensive. A physical bound on $\gamma_{max}$ is specified as a log-uniform distribution from $\gamma_{inj}$ to $10^3\gamma_{inj}$ which allows for cutoffs far above the Fermi window. Finally, a weakly informative normal distribution is used as the prior for p centered at p=3. This informative prior is set by considering the typical high-energy photon energies observed by Fermi and in most cases the data are more informative than the prior.

The GBM detectors are calibrated with an uncertainty in the total effective area of ~10%[34]. Therefore, an effective area correction, scaled to the brightest NaI detector, is allowed to vary for each detector within a normal prior centered at unity with a standard deviation of 0.1. These constants are allowed to vary in each spectral fit[21]. All spectral analysis and data reduction is performed within the 3ML framework[35]. Our radiative code is implemented in C++ and interfaced via Cython into astromodels[36].

**Synchrotron Modeling**

The evolution of the electron distribution function under synchrotron cooling alone obeys a Fokker-Planck type equation:

$$\frac{\partial}{\partial t} n_e(\gamma, t) = \frac{\partial}{\partial \gamma} C(\gamma) n_e(\gamma, t) + Q(\gamma)$$

where t is comoving time. Herein, we assume that the electrons are injected continuously between Lorentz factor $\gamma_{inj}$ and $\gamma_{max}$ with a constant spectral index $p$ such that

$$Q(\gamma) \propto \gamma^{-p} ; \quad \gamma_{inj} \leq \gamma \leq \gamma_{max}$$

and are cooled via emission of synchrotron photons:

$$C(\gamma) = -\frac{\sigma_T B^2}{6 \pi m_e c} \gamma^2$$

Here, $\sigma_T$ is the Thomson cross-section, $m_e$ the mass of an electron and c the speed of light. $B$ is the magnetic field strength. We do not consider escape or any other cooling terms. Other cooling terms require assumptions about the geometry of the emitting region which may in fact be important, but will be addressed in future studies. Thus, we assume a spectral model that is very simplistic.

The initial conditions for the cooling solution are setup in the following manner. The injected electron spectrum is set via $\gamma_{inj}$, $\gamma_{max}$ and p on a logarithmic grid of 300 points spanning from $\gamma = 1$ to $\gamma = \gamma_{max}$ The time step of the integration is set to the synchrotron cooling time of electrons with Lorentz factor $\gamma_{max}$ in order to resolve the cooling of the highest energy electrons. The number of time steps is computed via setting $\gamma_{cool}$ and calculating the time it takes for an electron at $\gamma_{max}$ to reach $\gamma_{cool}$ via synchrotron losses. In order to numerically solve the cooling equation, we employ the implicit Chang and Cooper method[37], which is computationally cheap and unconditionally stable.

At each time step, the electron distribution is convolved with the single-particle synchrotron emissivity:

$$n_\gamma(E) = \int_1^{\gamma_{max}} d\gamma \, n_e(\gamma, t) \, \Phi(E/E_{crit}(\gamma; B))$$

where

$$\Phi(w) = \int_w^\infty dx \, K_{5/3}(x)$$

$$E_{crit}(\gamma; B) = \frac{3}{2} \frac{B}{B_{crit}} \gamma^2$$

and the emission is summed with all previous emission resulting in the final photon spectrum. Geometric effects related to the jet geometry are not modeled but will only weakly influence the results and are highly model dependent. The solution to the cooling equation is calculated for each parameter set during the posterior sampling which is possible due to the fast numerical scheme employed.

The overall emission is characterized by five parameters: $B$, $\gamma_{inj}$, $\gamma_{cool}$, $\gamma_{max}$, and p. However, a strong degeneracy exists between $B$ and $\gamma_{inj}$ as their combination sets the peak of the photon spectrum via $\nu_{sync}(\gamma) = \Gamma B \, \gamma^2 q_e/2 \, \pi m_e c$. Thus, both parameters serve as an energy scaling which forces the setting of one of the parameters. We choose to set $\gamma_{inj} = 10^5$ though this choice is arbitrary. It is therefore important to note that all parameters are determined relatively, i.e., the values of $\gamma_{cool}$ and $\gamma_{max}$ are determined as ratios to $\gamma_{inj}$. Similarly, the value of $B$ is only meaningful when determining the characteristic energies of $\gamma_{cool}$ and $\gamma_{max}$ or $h\nu_{cool}$ and $h\nu_{max}$ respectively. In words, with our parameterization the spectra are scale free. The degeneracies can be eliminated by specifying temporal and radial properties of the GRB outflow which we will address in the next section.

**Model checking**

Assessing the viability of the spectral fits is performed under a Bayesian framework. Lacking other publicly available physical photon models for GRBs, model comparison is not possible for this study. Comparing to empirical models such as the Band function lacks statistical meaning as an empirical model can always be designed with more predictivity, but no physical basis. Therefore, we choose to model check our fits via posterior predictive checks (PPCs). One of the most natural ways to assess models via PPCs is graphically[38]. For all of our spectral fits, we simulate 500 replicated spectra from the posterior and compare these data to the observations in folded count space for each detector. Then, the fits are visually inspected for strong outliers, i.e., data points which are not predicted with the 95% credible region of the replicated data.

Admittedly, visual qualification of spectral fits is unsatisfying. However, this is currently the state-of-the-art in the statistical literature and is much more powerful than bootstrap methods which ignore the probability density of the posterior.

The posterior predictive distribution is the probability of replicated data given the observationally conditioned posterior. Mathematically,

$$p(\tilde{y}\,|\,y) = \int d\theta\, p(\tilde{y}\,|\,\theta) p(\theta\,|\,y)$$

where $y$ is our data, $\theta$ is our posterior and $\tilde{y}$ are replicated data from the conditioned model. In words, the distribution assesses the ability of the conditioned model to predict future data. If, for example, an instrument's response function was poorly modeled, replications from the posterior would systematically fail to predict these unmodeled features in the observed data. In order to check for discrepancies between fitted model and data, we compute replicated counts data from the posterior of our fits and construct quantile-quantile (QQ) plots[39,40] by computing the cumulative net count rate of the observed and replicated data. The 65% and 95% percentiles from 500 replications are plotted. We consider a fit to be poor when all NaI detectors deviate from the one-to-one line over a significant portion of the 95% region (see Extended Data Figs. 10-12).

**Physical Calculations**

From the results of the fits, we seek to estimate the physical parameters characterizing the emission region. We begin by assuming a radius and calculating an estimate of the bulk Lorentz factor via the relation $\Gamma = \sqrt{R/(2c\,t_p)}$ where $t_p$ is the duration of the pulse. This is equivalent to assuming that the time during which the electrons cool is equal to the expansion time. Henceforth, a quantity $\tilde{Q}$ implies a quantity computed after disentangling the scaled parameters from the fit, i.e., while we fit for $B$ which is degenerate with $\gamma_{inj}$, we will use these assumptions to compute $\tilde{B}$ via the posterior. Using our measured values of $h\nu_{inj}$ and $h\nu_{cool}$, we can now disentangle both $B$ and $\gamma_{inj}$ via

$$\tilde{B} = \left(\frac{18\pi q_e m_e c}{\Gamma \nu_{cool} t_p^2 \sigma_T^2}\right)^{1/3}$$

$$\tilde{\gamma}_{inj} = \sqrt{\frac{\nu_{inj} 2\pi m_e c}{q_e \Gamma \tilde{B}}}$$

where $q_e$ is the electron charge. To compute an estimate of the number of radiating electrons, we use the fact that most of the electrons are at $\gamma_\star = min(\tilde{\gamma}_{inj}, \tilde{\gamma}_{cool})$, and that the corresponding peak spectral flux is

$$F_\nu^{max} = \frac{N_e P_{\nu^{max}}}{4\pi d_L^2}$$

Where $P_{\nu^{max}}$ is the differential power at $\gamma_\star$ resulting in

$$N_e = \frac{3 q_e F_\nu^{max}}{m_e c^2 \sigma_T \Gamma \tilde{B} 4\pi d_L^2}.$$

With this formulation, we can compute $N_e$ in each time-interval of each GRB and sum to estimate the number of emitted electrons.

In the framework of the thermally accelerated fireball model, the characterization of the photospheric emission component fully characterizes the properties of the flow[41]. Our fits of optically-thin synchrotron radiation did not require the addition of a photospheric component. Further assuming that the energy radiated at the photosphere is smaller than the energy radiated in the optically thin region (which is a very loose assumption, since the GBM energy band is fully dominated by the optically thin synchrotron emission), it is possible to compute an upper limit on the Lorentz factor. Assuming a jet launching radius ($r_0$) of $10^7$ cm, and a radiative efficiency of ~10% leads to

$$\Gamma_{obs} \leq \Gamma_{photosphere} = \frac{\sqrt{2}}{4} \frac{\varepsilon^{1/8} L_{obs}^{1/4} \sigma_T^{1/4}}{\pi^{1/4} r_0^{1/4} (1+z)^{3/2} c^{3/4} m_p^{1/4}}$$

Where $L_{obs}$ is the total observed luminosity and $m_p$ is the proton mass. The lower limit is computed by requiring that the outflow is transparent to pair production above 1 MeV or

$$\tau_{\gamma\gamma} \simeq \frac{f_p \sigma_T L_{obs}}{\Gamma_{\gamma\gamma}^{4+2\beta} r_0^2 m_e c^2} < 1$$

where $f_p$ is the fraction of photons which can pair produce[17]. This implies

$$\Gamma_{obs} \geq \Gamma_{\gamma\gamma} = \left(\frac{f_p \sigma_T L_{obs}}{r_0^2 m_e c^2}\right)^{\frac{1}{4+2\beta}}$$

where β is the high-energy photon spectra index. Our inferred parameters imply that the upper limit $\Gamma_{photosphere}$ be smaller than the lower limit $\Gamma_{\gamma\gamma}$, which is a contradiction (Extended Data

Fig. 5). Thus we conclude that the standard thermal fireball model, together with optically thin synchrotron emission cannot explain these observations. Said differently, the entire prompt emission is required to be photospheric for the fireball model to be viable.

Motivated by the findings that the dissipation mechanism is unlikely to be due to internal shocks, and that the jet is required to be magnetically dominated, we reparameterize the problem as a function of the magnetization $\xi_B = U_B/U_e$ where $U_B = B^2/(8\pi)$ is the magnetic energy density and $U_e$ is the energy density in the heated electrons. Instead of imposing a radius, we infer radii for the entire sample assuming values of $\xi_B$ ranging from $10^{-2} - 1$. In addition, we allow for comoving bulk motion of the emission sites (blobs or mini-jets) to have Lorentz factors $\Gamma_{em}$ ranging from $1 - 10$ in the comoving frame[17,42].
Reparameterizing and inverting the previous relations yields

$$\Gamma = 1.8 \times 10^2\, \xi_{B,-1}^{3/16}\, \nu_{cool,MeV}^{7/32}\, d_{L,28}^{3/8}\, \nu_{inj,MeV}^{3/32}\, F_{\nu,26}^{max\ 3/16}\, \Gamma_{em,1}^{-9/8}\, t_{p,2s}^{-1/8}$$

$$R = 3.9 \times 10^{15}\, \xi_{B,-1}^{3/8}\, \nu_{cool,MeV}^{7/16}\, d_{L,28}^{3/4}\, \nu_{inj,MeV}^{3/16}\, F_{\nu,26}^{max\ 3/8}\, \Gamma_{em,1}^{-9/4}\, t_{p,2s}^{3/4}\ \mathrm{cm}$$

$$N = 2.74 \times 10^{48}\, \xi_{B,-1}^{-3/8}\, \nu_{cool,MeV}^{3/16}\, d_{L,28}^{7/4}\, \nu_{inj,MeV}^{-1/16}\, F_{\nu,26}^{max\ 7/8}\, \Gamma_{em,1}^{-5/4}\, t_{p,2s}^{3/4}$$

$$B = 0.3\, \xi_{B,-1}^{-1/16}\, \nu_{cool,MeV}^{-13/32}\, d_{L,28}^{-1/8}\, \nu_{inj,MeV}^{-1/32}\, F_{\nu,26}^{max\ -1/16}\, \Gamma_{em,1}^{11/8}\, t_{p,2s}^{-5/8}\ \mathrm{G}$$

$$\gamma_{min} = 8.4 \times 10^4\, \xi_{B,-1}^{-1/16}\, \nu_{cool,MeV}^{3/32}\, d_{L,28}^{-1/8}\, \nu_{inj,MeV}^{15/32}\, F_{\nu,26}^{max\ -1/16}\, \Gamma_{em,1}^{-5/8}\, t_{p,2s}^{3/8}$$

Specifically for the radius, we see that the dependence on the parameters is $\xi_{B,-1}^{3/8}\Gamma_{em,1}^{-9/4}$. Lowering the radius to be smaller than the deceleration radius requires us to decrease $\xi_B$ by several order of magnitude or increase $\Gamma_{em}$ by a few. Indeed, emission radii are required to be high, but not so high that they violate the deceleration radius. Thus these models have been constrained to a tight parameter space.

Finally, we note that in the above derivations time is linked via

$$t_{dyn} = \frac{R}{\Gamma \Gamma_{em} c} \equiv t_{sync} = \frac{3 m_e c}{4 \sigma_T U_B \gamma_c^2}\ .$$

In words, the cooling time of the electron is associated to the dynamical time expressed in the blob/mini-jet frame. This breaks the dependence between pulse duration and radiative time by the introduction of an additional factor $\Gamma_{em}$.

**Extended Data Figure 1: Evolution of parameters**
For GRBs with several time bins, we display the evolution of the luminosity (red; right y-axis), $h\nu_{cool}$ and $h\nu_{inj}$ (green/blue; left y-axis), respectively. No clear, common trend is observed except that there appears to be a transition between slow and fast cooling during the decay of the luminosity. The GBM spectral window is displayed as the grey region.

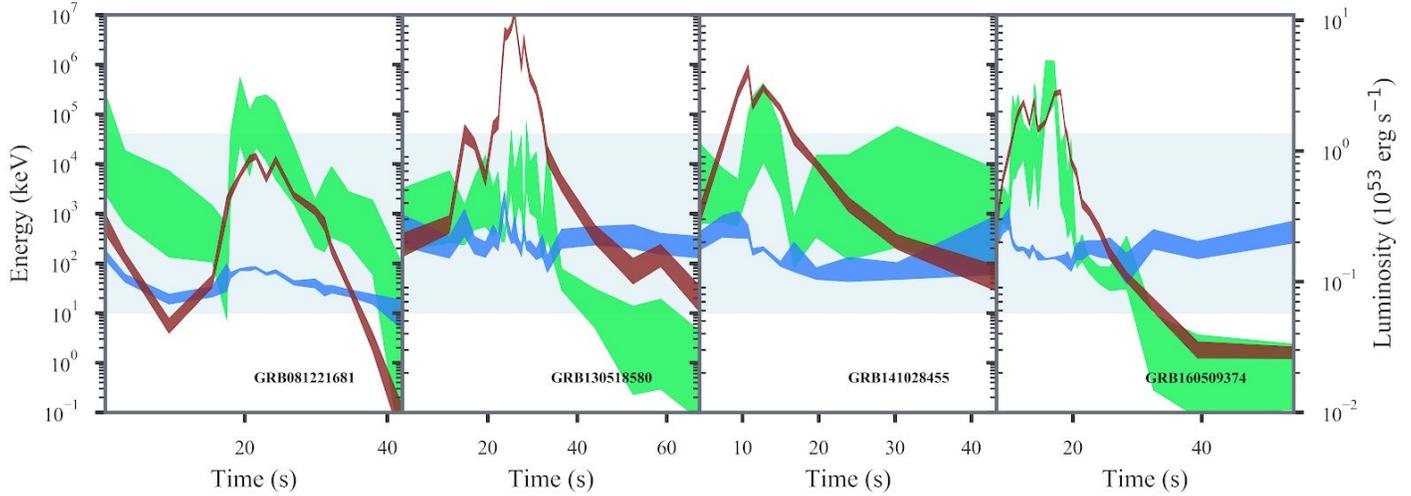

**Extended Data Figure 2: Counting electrons**
The inferred number of emitting electrons assuming an emission radius of $10^{14}$ cm under the fireball framework. The number is derived by computing the observed luminosity from our synchrotron modeling and comparing it to the predicted luminosity of synchrotron emission. The ratio of the two quantities yields an estimate the number of emitting electrons.

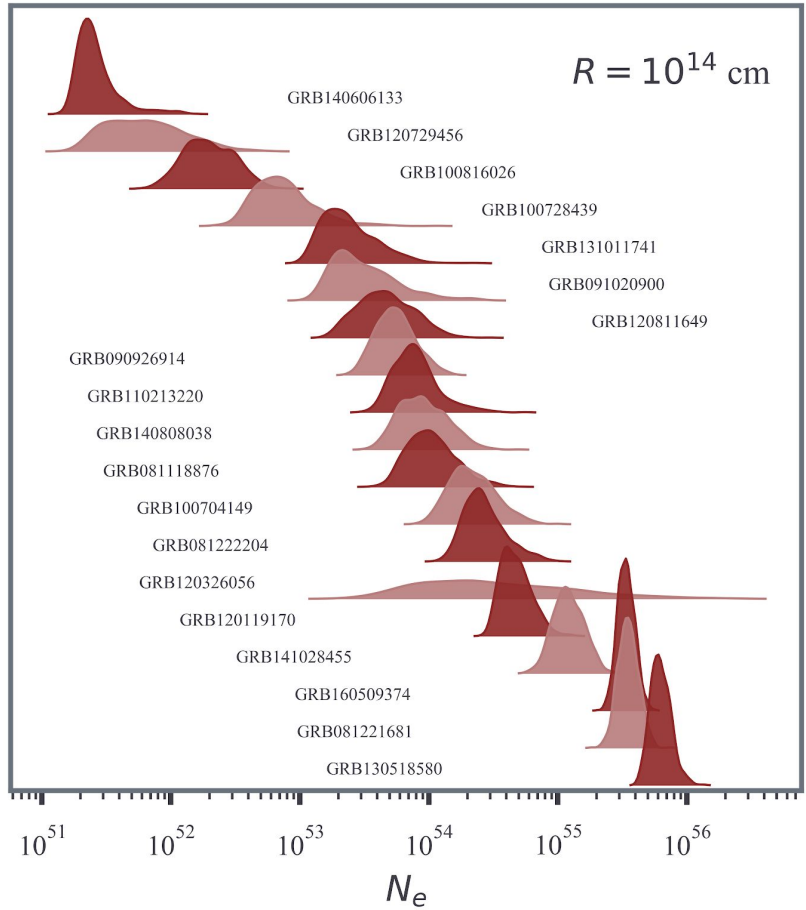

**Extended Data Figure 3: The inferred magnetic field**

By assuming an emission radius, the magnetic field (B) can be disentangled from the minimum/injection electron Lorentz factor $\gamma_{inj}$ and we can examine its correspondence with the cooling regime. Interestingly, higher magnetic field strength strongly correlates with faster-cooling, as predicted. This correlation is not completely set by our parameterization, but inter-burst variation of the emission radius can alter the relative values of B.

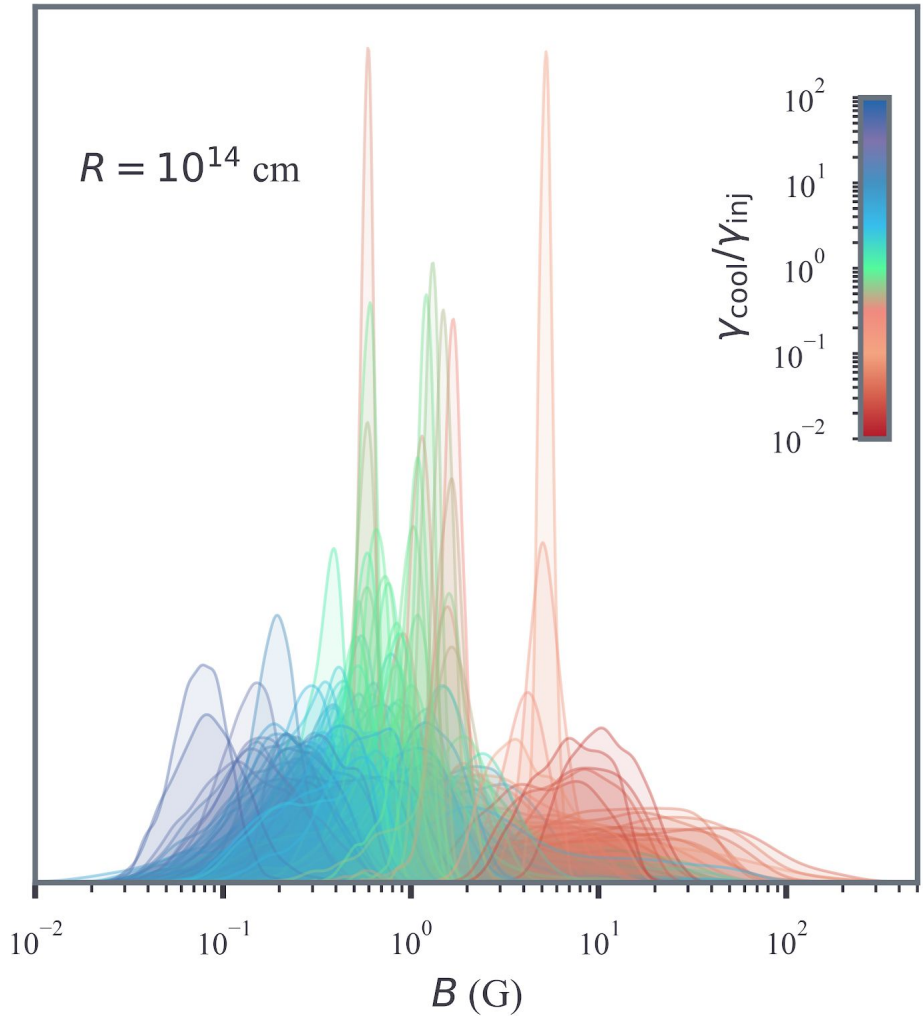

**Extended Data Figure 4: The electron distribution spectral index**
The distribution of the injected electron spectral indices is nearly constant as a function of cooling. There is a slight trend to steeper indices when there is little cooling. The inclusion of a cutoff in our model should compensate for an erroneously steep power law, thus the very steep values we find are likely not an artifact of sensitivity. Both shock accelerations and magnetic reconnection can be reconciled with the few steep power laws we observe, either with very oblique shock geometry or low magnetization, respectively.

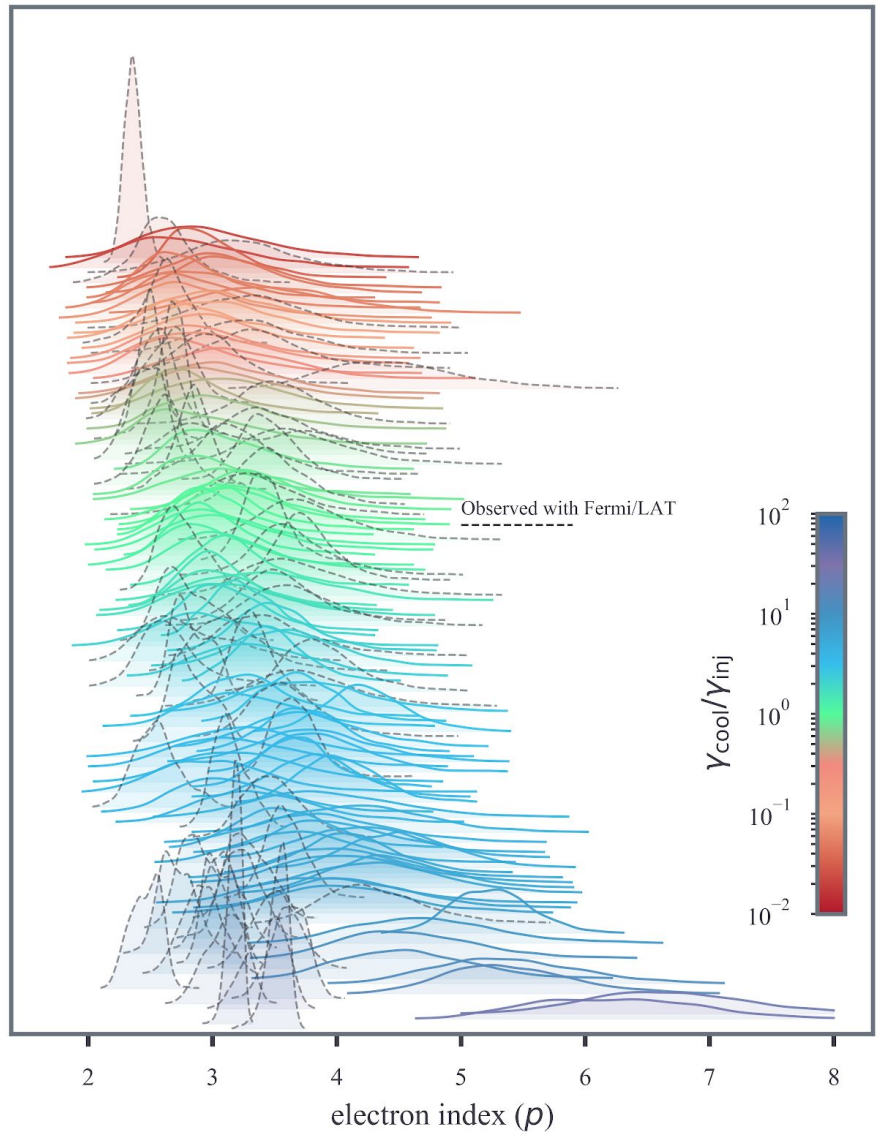

**Extended Data Figure 5: A contradiction in the thermal fireball**
The upper and lower limits for each GRB outflow's bulk Lorentz factor derived from the absence/subdominance of a photospheric component and the assumption that the flow (in the rest frame) must be transparent to gamma-rays that would otherwise produce electron/positron pairs above 1 MeV. Both limits are subject to assumptions, with weak dependence on the conclusions.

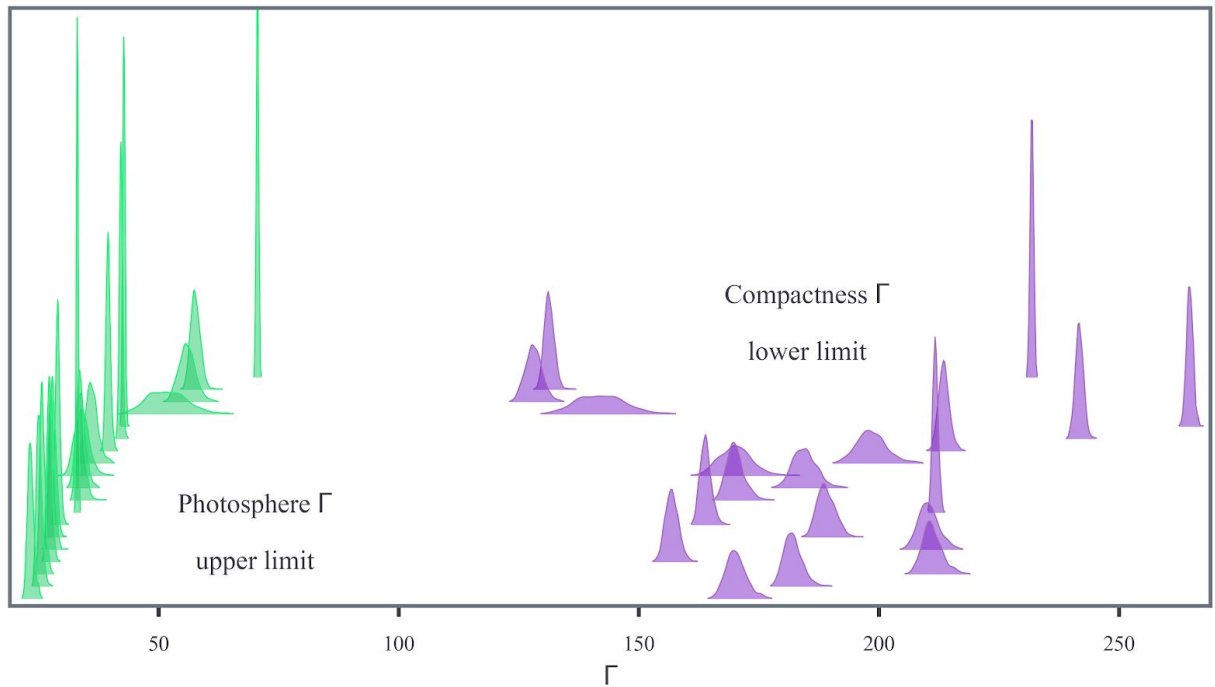

**Extended Data Figure 6: Radii inferred from magnetization and mini-jets**
By parameterizing the jet in terms of the fractional magnetization ($\xi_B$) and allowing for comoving bulk motion ($\Gamma_{em}$), we compute the inferred radii for all spectra in our sample.

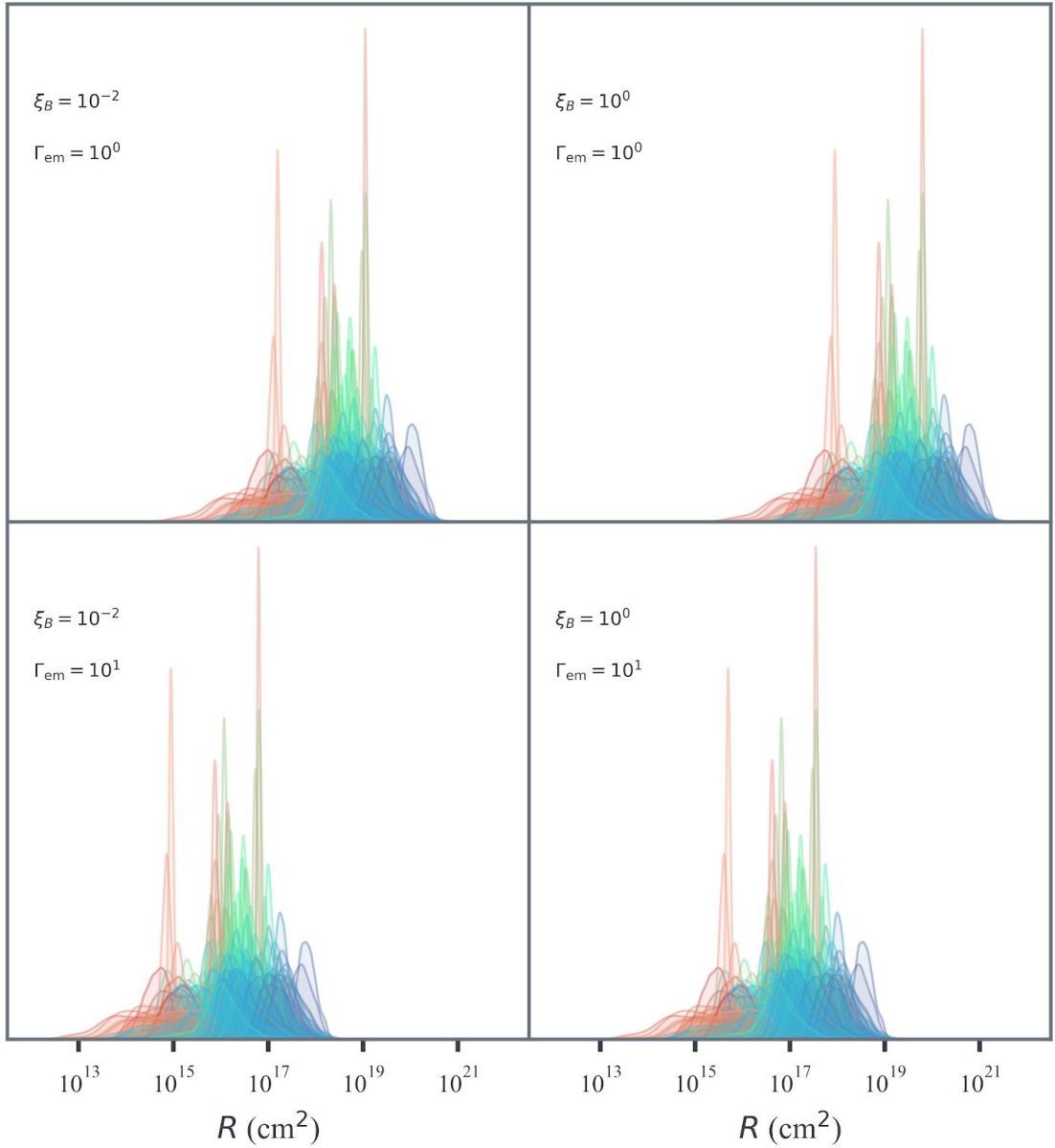

**Extended Data Figure 7: Bulk Lorentz factors inferred from magnetization**
By parameterizing the jet in terms of the fractional magnetization ($\xi_B$) and allowing for comoving bulk motion ($\Gamma_{em}$), we compute the inferred jet bulk Lorentz factors for all spectra in our sample.

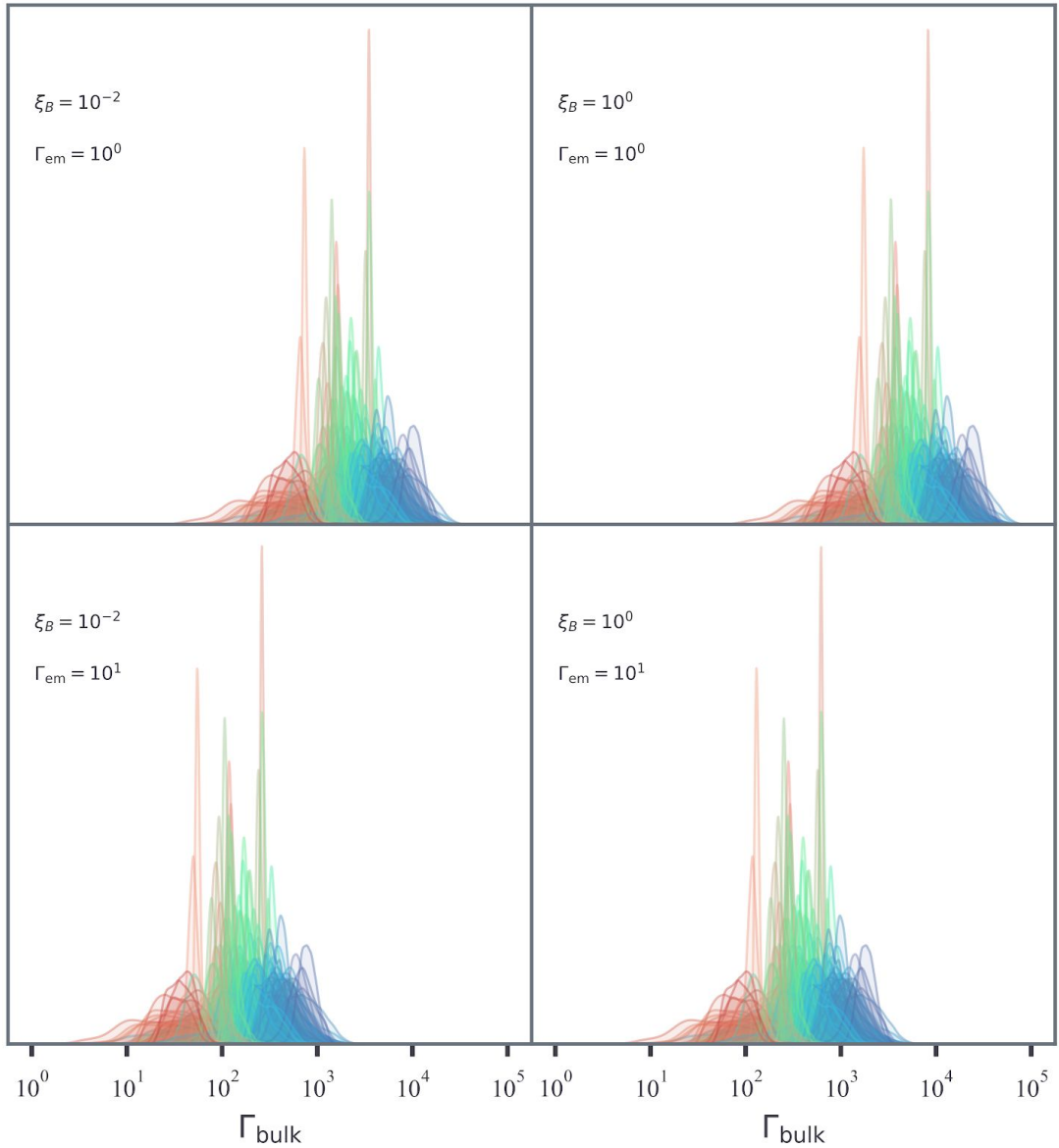

**Extended Data Figure 8: Relation between Zhang et al. (2016) and our model**

By refitting GRB 130606B with our synchrotron formulation, we can compare our fitted cooling regime to the fitted temporal injection index $q$ such that $n_e^{inj}(t) \propto t^q$ from [27]. We find a clear correlation indicating that the cooling regime in their model is controlled by the number of uncooled electrons injected at later times. Thus, that particular model [27] is strictly-speaking not a fast-cooling model.

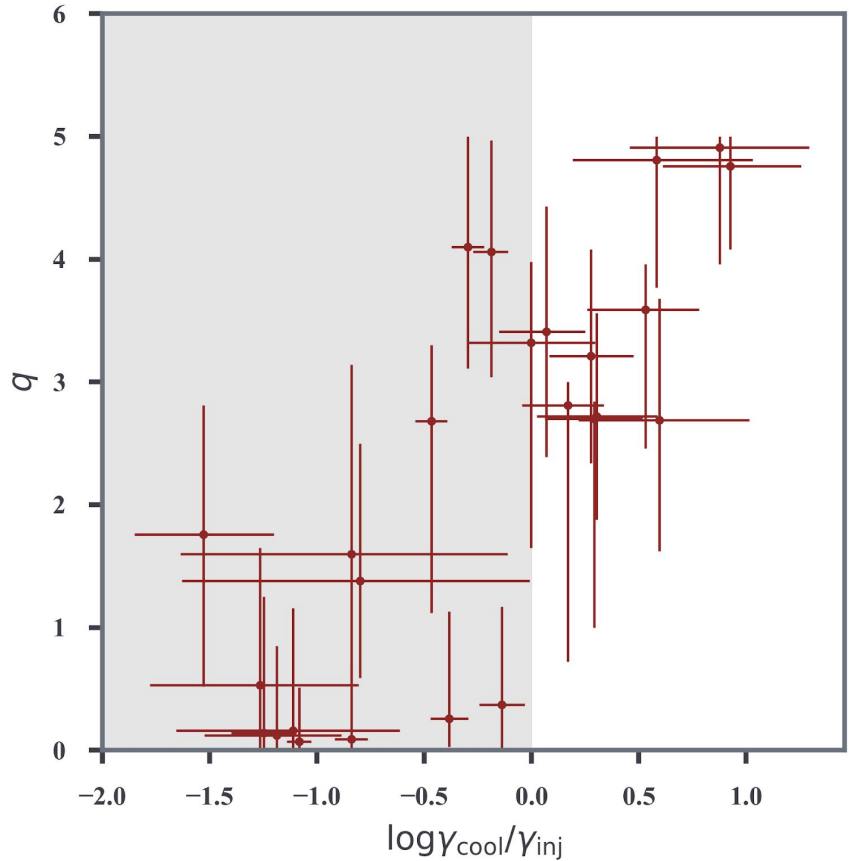

**Extended Data Figure 9: Electrons from decaying magnetic field**
The time-evolution of an electron distribution when considering a decaying magnetic field and increasing injection with time according to [27]. The initially injected electrons are cooled but quickly dominated by uncooled electrons injected at later times. Thus, the model [27] can produce spectral shapes similar to the modeling used in this work, yet, with more fine-tuned assumptions. However, the efficiency problems of synchrotron emission are not alleviated with this formalism.

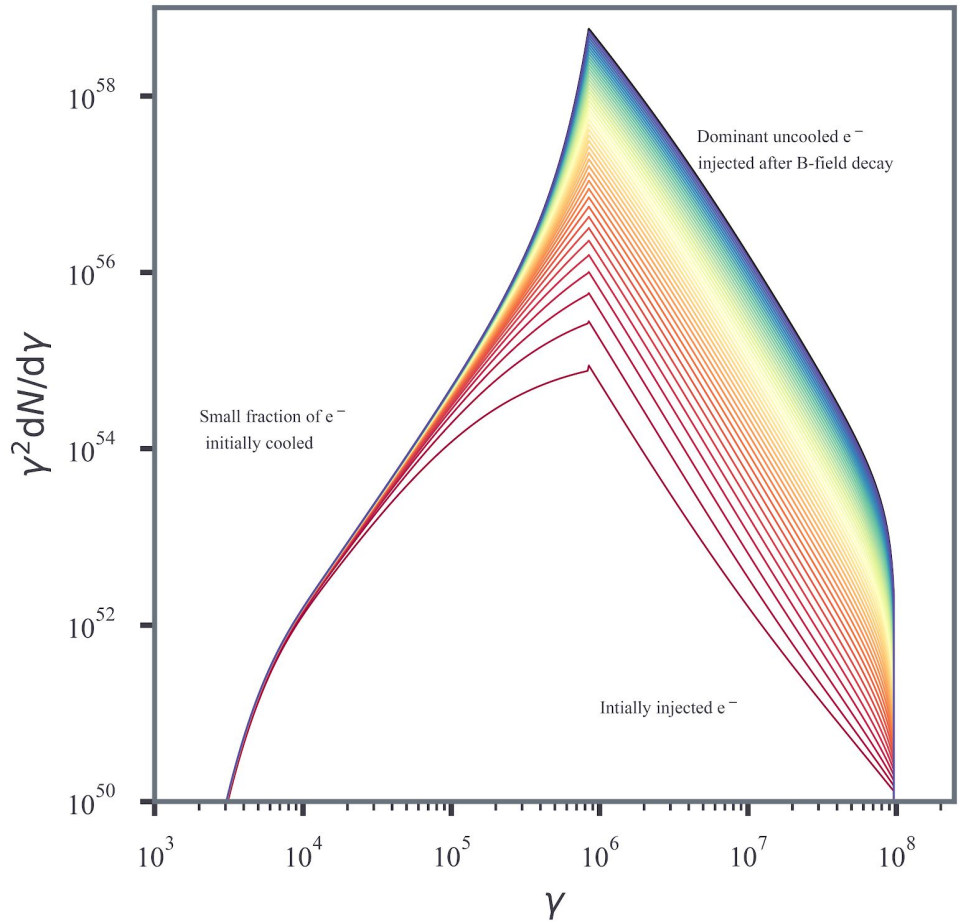

**Extended Data Figure 10: PPC-QQ plot of a good fit**

500 realizations of the posterior predictive distribution for each detector are created and the cumulative synthetic data are compared to the cumulative real data for every realization. The 68% and 95% quantiles are displayed along with the one-to-one relation (green) which indicates a perfect fit. While there is some structure in the curves, the green line is within the synthetic and resembles what we would expect from a simulated synchrotron observation (see Extended data Figure 12).

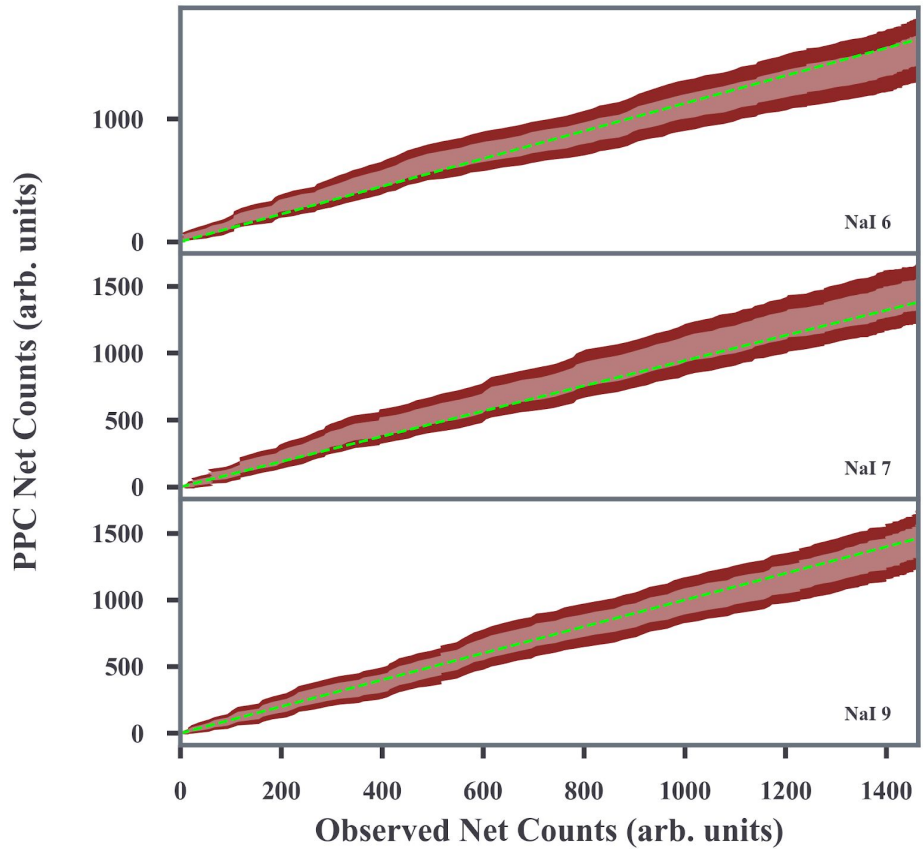

**Extended Data Figure 11: PPC-QQ plot of a bad fit**
An example of a poorly fit spectrum in our sample. The highlighted regions indicate where the replicated model deviates strongly from the observed data.

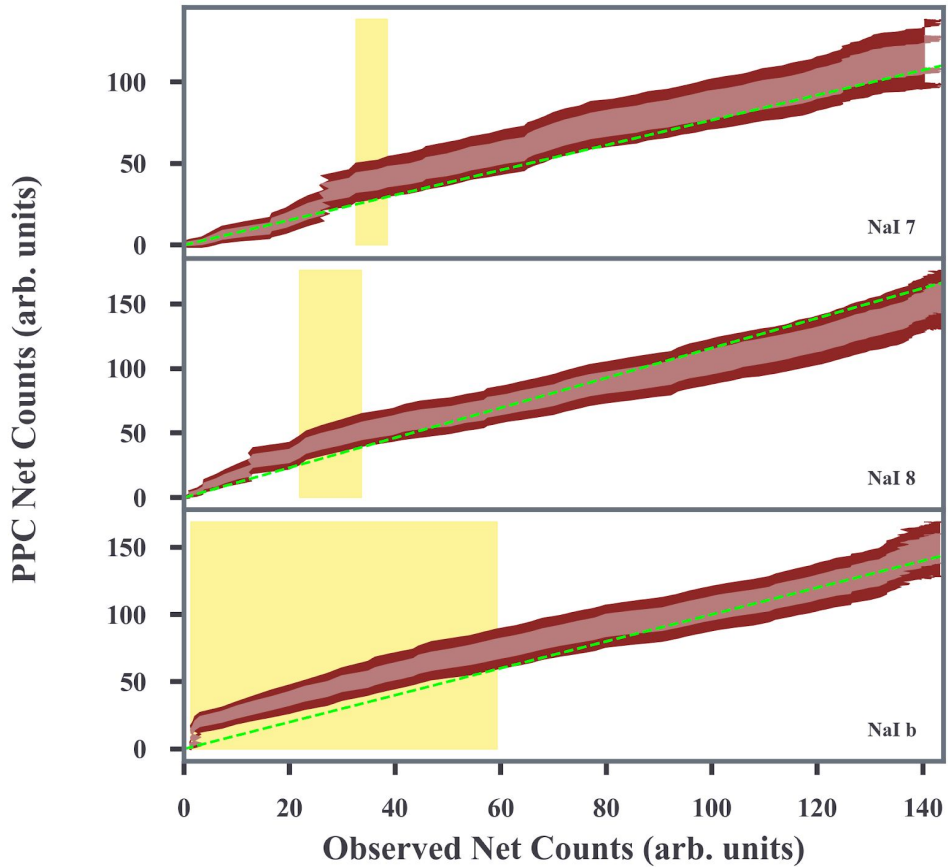

**Extended Data Figure 12: PPC-QQ plot of a simulated synchrotron spectrum**
We simulate a synthetic synchrotron spectrum and compute the PPC-QQ plot from the fit. Thus we know that the true model is synchrotron. While there are no strong deviations from the one-to-one relation, there does exist structure in the curve reflecting the variance in the data. This demonstrates the danger of using frequentist point-estimate residuals to judge the quality of a fit.

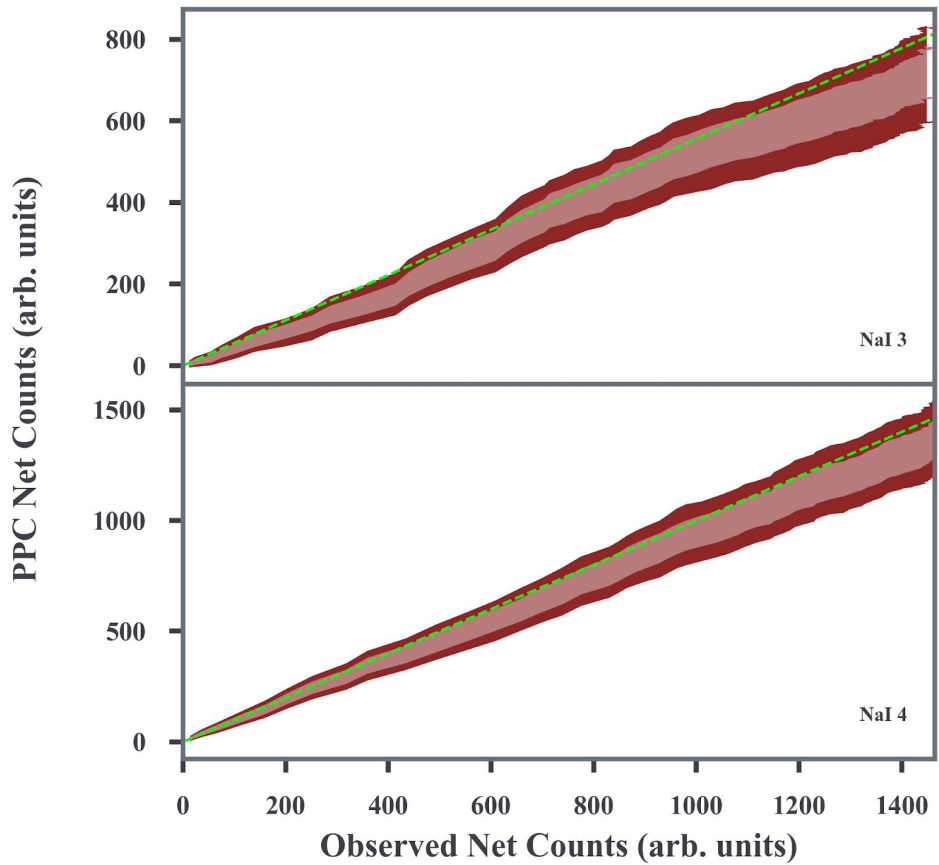